\documentclass[superscriptaddress,twocolumn,floatfix,aps,reprint,preprintnumbers]{revtex4-1}

\usepackage{soul}
\usepackage{textcomp}
\usepackage{graphicx}
\usepackage{amsmath}
\usepackage{xcolor}
\usepackage{bm}
\usepackage[hidelinks]{hyperref}

\begin{document}

\title{Evidence of superfluidity in a dipolar supersolid from non-classical rotational inertia}
\author{L. Tanzi}
\affiliation{CNR-INO, S.S. A.~Gozzini di Pisa, via Moruzzi 1, 56124 Pisa, Italy}
\affiliation{LENS and Dip. di Fisica e Astronomia, Universit$\grave{a}$ di Firenze, 50019 Sesto Fiorentino, Italy}
\author{J. G. Maloberti}
\affiliation{CNR-INO, S.S. A.~Gozzini di Pisa, via Moruzzi 1, 56124 Pisa, Italy}
\affiliation{LENS and Dip. di Fisica e Astronomia, Universit$\grave{a}$ di Firenze, 50019 Sesto Fiorentino, Italy}
\author{G. Biagioni}
\affiliation{CNR-INO, S.S. A.~Gozzini di Pisa, via Moruzzi 1, 56124 Pisa, Italy}
\affiliation{LENS and Dip. di Fisica e Astronomia, Universit$\grave{a}$ di Firenze, 50019 Sesto Fiorentino, Italy}
\author{A. Fioretti}
\affiliation{CNR-INO, S.S. A.~Gozzini di Pisa, via Moruzzi 1, 56124 Pisa, Italy}
\author{C. Gabbanini}
\affiliation{CNR-INO, S.S. A.~Gozzini di Pisa, via Moruzzi 1, 56124 Pisa, Italy}
\author{G. Modugno}
\affiliation{CNR-INO, S.S. A.~Gozzini di Pisa, via Moruzzi 1, 56124 Pisa, Italy}
\affiliation{LENS and Dip. di Fisica e Astronomia, Universit$\grave{a}$ di Firenze, 50019 Sesto Fiorentino, Italy}
\email{Electronic address: modugno@lens.unifi.it\\}

\begin{abstract}
A key manifestation of superfluidity in liquids and gases is a reduction of the moment of inertia under slow rotations. Non-classical rotational effects have also been considered in the context of the elusive supersolid phase of matter, in which superfluidity coexists with a lattice structure. Here we show that the recently discovered supersolid phase in dipolar quantum gases features a reduced moment of inertia. Using a dipolar gas of dysprosium atoms, we study a peculiar rotational oscillation mode in a harmonic potential, the scissors mode, previously investigated in ordinary superfluids. From the measured moment of inertia, we deduce a superfluid fraction that is different from zero and of order of unity, providing direct evidence of the superfluid nature of the dipolar supersolid.
\end{abstract}

\date{\today}
\maketitle

Superfluids exhibit their most spectacular properties during rotation. This is because the superfluid state is described by a macroscopic wavefunction, whose phase can change only by integer multiples of  2$\pi$ upon completing a closed path. For a cylindrical superfluid rotating at low angular velocities, $\omega\to0$, this condition leads to the vanishing of both angular momentum $L$ and moment of inertia $I=\langle L\rangle/\omega$. An angular momentum can appear only for sufficiently large $\omega$ at integer multiples of the reduced Planck’s constant $\hbar$, through the appearance of quantized vortices. These non-classical rotational effects have been verified for most known superfluids: nuclear matter~\cite{0}, $^4$He~\cite{1}, $^3$He~\cite{2}, gaseous Bose-Einstein condensates~\cite{3}, degenerate Fermi gases~\cite{4} and exciton-polariton condensates~\cite{5}. A related phenomenon is the Meissner effect in superconductors~\cite{6}.

At the end of the ’60, another type of bosonic phase of matter described by a macroscopic wavefunction, the supersolid, was predicted to exist. In a supersolid, superfluidity coexists with a crystal-type structure~\cite{7,8,9}. A. J. Leggett suggested that a rotating supersolid should show a moment of inertia intermediate between those of a superfluid and of a classical system, $I=(1-f_s)I_c$. Here, $I_c$ is the classical moment of inertia and $0\leq f_s\leq1$ is the so-called superfluid fraction~\cite{9}. This phenomenon is called non-classical rotational inertia (NCRI). Standard superfluids can have $f_s<1$, but only at finite temperature, $T>0$, thanks to the presence of a thermal component. In a supersolid at $T=0$, the reduction of the superfluid fraction is instead caused by the spatially modulated density, which tends to increase the inertia towards the classical limit~\cite{9,10}.

At the time it was proposed, the primary candidate for observing supersolidity was solid helium. Torsion oscillators were employed extensively to attempt detecting NCRI~\cite{11}. The original announcement of the possible presence of a large superfluid fraction, $f_s\approx10^{-1}$~\cite{12,13}, has later received a different interpretation based on a change of the elastic properties of the solid~\cite{14} and has not been confirmed by more recent studies~\cite{15}. Superfluidity in bulk solid helium has now been excluded down to the level of $10^{-4}$~\cite{16}, and the search goes on in 2D films~\cite{17}.

Here, we study a different supersolid candidate, a gaseous Bose-Einstein condensate (BEC) of strongly dipolar atoms, where a density-modulated regime coexisting with the phase coherence necessary for supersolidity has been recently discovered~\cite{18,19,20}. So far, its superfluid nature has been tested through non-rotational excitation modes that can be described in terms of the hydrodynamic equations for superfluids~\cite{21,22,23}. Here we aim instead at characterizing the NCRI of such system, searching for direct evidence of superfluidity under rotation, in the spirit of the helium experiments.

Achieving dipolar supersolids large enough to realize a cylindrical geometry is so far not possible, so we employ a specific rotation technique that fits the asymmetric, small-sized systems available in the laboratory. We excite the so-called scissors mode, a small-angle rotational oscillation of  the harmonic potential that naturally holds the system. This technique, inspired by an excitation mode of nuclei~\cite{26}, has been proposed~\cite{24,25} and employed~\cite{27} to demonstrate superfluidity of ordinary BECs. A recent theoretical study has shown that the scissors mode can also be used to characterize the NCRI of a dipolar supersolid~\cite{28}. We study the change of the scissors mode frequency across the transition from BEC to the supersolid regime, so that we can directly compare the supersolid with a fully superfluid system. 

\begin{figure}[t]
\centering
\includegraphics[width=\columnwidth]{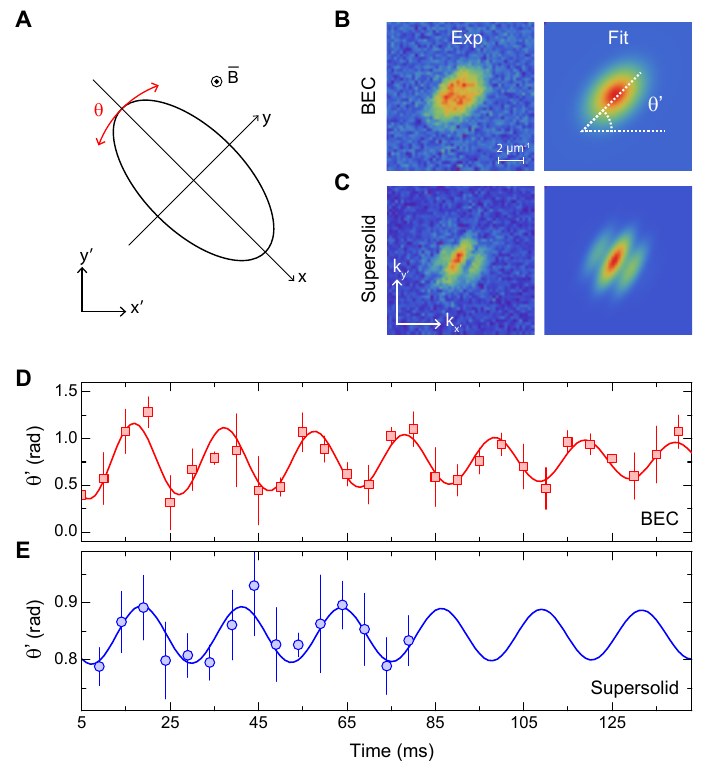}
\caption{Scissors mode measurements. A) Sketch of the experimental geometry: the atomic system (ellipse) is trapped in an anisotropic potential with eigenaxes $x$ and $y$. A sudden rotation of the trapping potential excites an angular oscillation $\theta(t)$ (red arrows). B-C) Examples of the experimental distributions after free expansion and of the corresponding two-dimensional fits used for extracting the oscillation angle $\theta'$ after the free expansion in B) BEC regime ($\epsilon_{dd}$=1.14); C) supersolid regime ($\epsilon_{dd}$=1.45). D-E). Time evolution of the angle $\theta'(t)$: D) BEC regime; E) supersolid regime. Error bars represent the standard deviation of 4-8 measurements.}
\label{fig1}
\end{figure}

In the experiment, a BEC of strongly magnetic Dy atoms is held in an anisotropic harmonic trap, with frequencies $\omega_{x,y,z}=2\pi$(23,46,90)~s$^{-1}$, with the dipoles oriented in the $z$ direction by a magnetic field $B$ (Fig.~\ref{fig1}). The temperature is sufficiently low to have a negligible thermal component \cite{29}. We induce the transition from BEC to supersolid by tuning via a magnetic Feshbach resonance the interaction parameter $\epsilon_{dd}$, which parametrizes the ratio of the dipolar and van der Waals interaction energies \cite{18}. In the supersolid regime, a density modulation develops along the weak $x$ axis, leading to the appearance of interference peaks in the momentum distribution. We expect our lattice to be composed of two principal density maxima, or “droplets”, each containing about $10^4$ atoms \cite{21}. This realizes a so-called cluster supersolid \cite{30}, very different from the hypothesized helium supersolid with one particle per lattice site. In principle, further tuning of $\epsilon_{dd}$ would bring the system into the so-called droplet crystal regime, with no coherence between the droplets \cite{18,19,20}.

The scissors mode is excited by changing suddenly the direction of the eigenaxes of the harmonic trap \cite{29}. This results in a sinusoidal oscillation with frequency $\omega_{sc}$ of the angle $\theta$ between the long axis of the system and the corresponding trap axis. We choose to rotate the system in the ($x,y$) plane, perpendicular to the direction of the dipoles, in order for the dipolar interaction potential to be independent of $\theta$~\cite{31,32}. 

The oscillation frequency can be directly related to the moment of inertia of the superfluid through:
\begin{equation}
I=I_c\,\alpha\beta\frac{\omega_x^2+\omega_y^2}{\omega_{sc}^2},
\label{eq1}
\end{equation}
where $\alpha=(\omega_y^2-\omega_x^2 )/(\omega_x^2+\omega_y^2)$ and $\beta=\langle x^2-y^2\rangle/\langle x^2+y^2\rangle$ are geometrical factors measuring the deviation from cylindrical symmetry of the trap and of the density distribution, respectively~\cite{24,28}. Whereas  $\alpha$ can be measured experimentally, $\beta$ needs to be calculated theoretically~\cite{29}. For non-dipolar BECs in the Thomas-Fermi regime, one has the simplification $\beta=\alpha$~\cite{25}. For dipolar systems, the density deformation changes instead with the interaction parameter owing to magnetostriction, $\beta=\beta(\epsilon_{dd})\neq\alpha$~\cite{31}. If the oscillation amplitude is much smaller than $\beta$, the density deformation stays constant during the motion~\cite{24}.

We can now connect the moment of inertia to a superfluid fraction, which we define specifically for our system in analogy with Leggett’s definition, taking into account our non-cylindrical geometry:
\begin{equation}
I=(1-f_s)I_c+f_s\beta^2I_c.
\label{eq2}
\end{equation}
It is easy to see that this definition coincides with Leggett’s one in the cylindrical case, $\beta=0$. It also coincides with the known results for a superfluid with elliptical geometry, $I=\beta^2 I_c$~\cite{0,24, 33}. The presence of a residual moment of inertia in the BEC, despite $f_s=1$ at $T=0$, derives from a peculiar velocity distribution, which is very different from the one in a cylindrical geometry~\cite{24,25}. Finally, by combining Eqs.~(\ref{eq1}) and eq.~(\ref{eq2}) one can directly relate the superfluid fraction to the trap and scissors frequencies and to the deformation:
\begin{equation}
f_s=\frac{1-\alpha\beta(\omega_x^2+\omega_y^2)/\omega_{sc}^2}{1-\beta^2}.
\label{eq3}
\end{equation}
We note that the scissors mode is analogous to the helium torsion oscillators because both detect NCRI via the oscillation frequency~\cite{12,13,14,15}, although there are some differences. In the scissors mode, all atoms experience the restoring force from the trap, so there are no elastic effects to consider~\cite{14}. A finite deformation $\beta$ is clearly necessary for the scissors mode, whereas torsion oscillators are normally symmetric, although macroscopic deformations can be taken into account with the same formalism~\cite{33,34} and a related tortuosity effect is present for superfluids in porous media~\cite{12}.

\begin{figure*}[htbp]
\centering
\includegraphics
{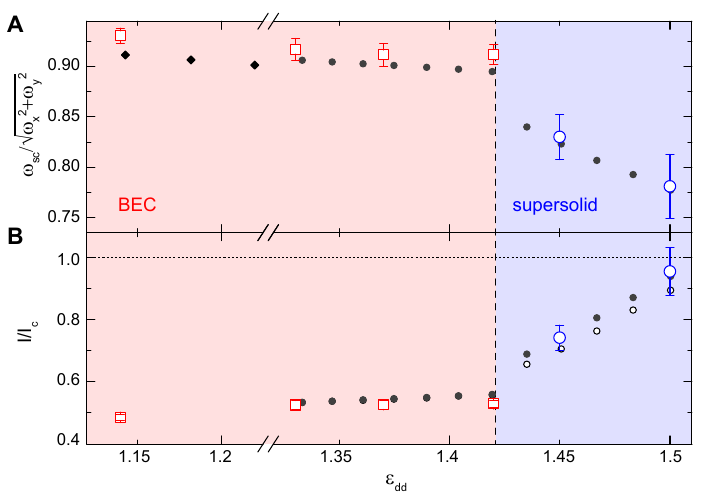}
\caption{Scissors mode frequency and moment of inertia vs the interaction parameter. A) Scissors mode frequencies. Large circles and squares are the experimental measurements. Black diamonds and dots are the mean-field and beyond-mean-field theoretical predictions, respectively~\cite{31,28}. B) Moment of inertia. Large squares and circles are derived from Eq.~(\ref{eq1}), using the experimental measurements of the scissors frequencies and the theoretical $\beta$~\cite{28}; black dots are the numerical simulation~\cite{28}. Small open dots are the theoretical prediction for $\beta^2$~\cite{28}. Error bars are one standard deviation ~\cite{29}. In the experiment, $\epsilon_{dd}$ has a calibration uncertainty of 3\%. The dashed line separating BEC and supersolid regimes was determined numerically~\cite{28}.}
\label{fig2}
\end{figure*}

Let us now turn to the experimental results. Figures~\ref{fig1}B-E, summarize the scissors measurements in the BEC and supersolid regimes. The 2D density distributions are imaged after a free expansion of the system, representing effective momentum distributions. They are fitted to extract the angle $\theta'$ in the laboratory frame for various observation times $t$. The resulting data for $\theta'$ are fitted with a sinusoid to measure $\omega_{sc}$ \cite{29}. Both BEC and supersolid regimes feature single-frequency oscillations, as expected for weakly-interacting superfluids \cite{24}. We have checked that a thermal sample features instead a two-frequency oscillation, as expected for a weakly interacting system~\cite{29}.

To avoid perturbations caused by other collective modes ~\cite{29}, we employ two different excitation techniques for the BEC and the supersolid regimes, which result in a lower amplitude of the scissors mode for the supersolid, see Figs.~\ref{fig1}D-E. The accuracy in the determination of the scissors frequency in that regime is limited also by the finite lifetime of the supersolid~\cite{18}. 

A summary of the experimental results for the scissors frequency and the related moment of inertia is shown in Fig.~\ref{fig2}. The results are compared to the theoretical predictions of Ref.~\cite{28}, calculated for trap parameters and atom numbers close to the experimental ones. For the BEC, we measure a frequency that depends only weakly on the interaction parameter $\epsilon_{dd}$, consistently with the prediction of a weak change of the deformation $\beta(\epsilon_{dd})$~\cite{31}. In contrast, when the system enters the supersolid regime we observe a clear reduction of the frequency, in agreement with the theory. From the measured frequency, we can determine the moment of inertia $I/I_c$, through Eq.~(\ref{eq1}), where the deformation $\beta$ is determined from the numerically calculated density distributions~\cite{28}. The results are shown in Fig.~\ref{fig2}B. In the BEC regime, the moment of inertia differs by a factor of two from the classical value and is compatible with $\beta^2$, as expected for a fully superfluid system. In the supersolid regime, at $\epsilon_{dd}$=1.45, the moment of inertia increases towards the classical value, but does not reach it. This provides evidence of NCRI for the dipolar supersolid.

The data point in Fig.~\ref{fig2}B further in the supersolid regime, at $\epsilon_{dd}$=1.5 has larger error bars owing to the shorter lifetime of the system. We were unable to study the droplet crystal regime, which is predicted to appear for $\epsilon_{dd}\approx$1.52~\cite{28}, because of the loss of the interference pattern~\cite{18,19,20}.

\begin{figure*}[t]
\centering
\includegraphics
{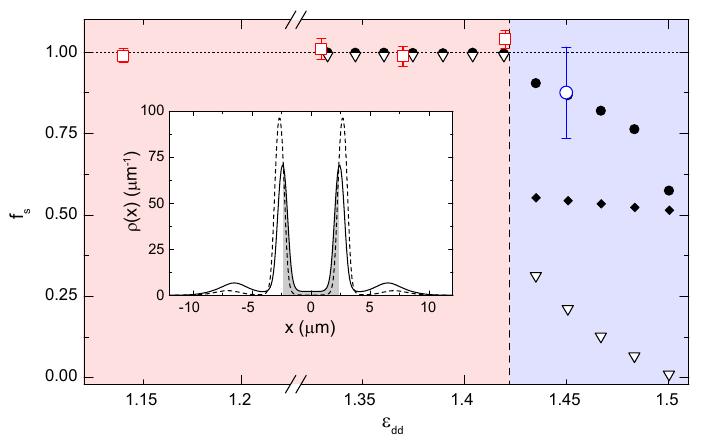}
\caption{Superfluid fraction from BEC to supersolid. Red squares and blue circles are the superfluid fraction from the experimentally measured scissors frequency and the theoretical $\beta$, using Eq.~(\ref{eq3}). Black dots are the superfluid fraction from the theoretical frequency \cite{28}. Open triangles are the upper limit for the one-dimensional superfluid fraction from Eq.~\ref{eq4}. Diamonds are the estimated superfluid fraction of independent droplets. Inset: calculated mean density distribution \cite{29} for $\epsilon_{dd}$=1.45 (continuous line) and $\epsilon_{dd}$=1.5 (dashed line); the gray region is the region of integration for Eq.~\ref{eq4}.}
\label{fig3}
\end{figure*}

The change of $I/I_c$ is in principle due to both the change of shape, $\beta(\epsilon_{dd})$, when the supersolid modulation forms and the related change of the superfluid fraction. The experiment-theory agreement for $I/I_c$ both in the BEC regime, where $f_s$=1, and at $\epsilon_{dd}$=1.5, where $I$ is expected to be close to $I_c$, supports the validity of the calculated $\beta$ for our system. Equation~\ref{eq2} shows that if the superfluid fraction of the supersolid varies between 0 and 1, then $I/I_c$ shown in Fig.~\ref{fig2}B should vary between 1 and $\beta^2$. More directly, we calculate the superfluid fraction from Eq.~\ref{eq3}, employing the experimental frequencies and the theoretical $\beta$. The results are shown in Fig.~\ref{fig3}, together with the corresponding points calculated from the theoretical predictions of ref.~\cite{28}.

In the BEC regime, the data confirm that the system is fully superfluid, $f_s$=1, as already found for non-dipolar BECs \cite{27}. In the supersolid regime, we can reliably calculate the superfluid fraction only for the experimental data point at $\epsilon_{dd}$=1.45. Remarkably, the superfluid fraction of the supersolid is very large, $f_s\approx$0.9, in agreement with the numerical calculations. Given the measurement uncertainty, $f_s$ is consistent with unity and inconsistent with zero. This result demonstrates the superfluid nature of the dipolar supersolid under rotation. 

The theory predicts a reduction of the superfluid fraction moving further into the supersolid regime, although $f_s$ remains finite even in the droplet crystal regime because of the superfluidity of the individual droplets under rotation~\cite{28}. In the experiment, we cannot check whether $f_s$ decreases moving to $\epsilon_{dd}$=1.5, because the lower measurement accuracy and the increase of $\beta^2$ shown in Fig.~\ref{fig2}B prevent us from measuring $f_s$ reliably \cite{29}.

It is interesting to compare our results to the original prediction by Leggett for the superfluid fraction of a supersolid rotating in a one-dimensional annulus,
\begin{equation}
f_s\leq\left(\int\frac{dx}{\rho(x)}\right)^{-1}.
\label{eq4}
\end{equation}
where $\rho(x)$ is the normalized density along the annulus and the integral is performed on a lattice cell~\cite{9,10}. Equation~\ref{eq4} shows that the reduction of the superfluid fraction is a consequence of the breaking of translational invariance, as $f_s$ is determined by the minimum density between lattice sites. Intuitively, in a homogenous superfluid, $\rho(x)$=const implies that each atom is equally delocalized so no rotation happens. In a system where $\rho(x)\to0$ between neighboring lattice sites, the sites are distinguishable so the system rotates classically. The supersolid is the intermediate case in which the atoms are still delocalized but the density modulation allows a partial rotation, increasing the moment of inertia when compared to a homogeneous superfluid. 

In 1970, Leggett used Eq.~\ref{eq4} and the known information on the helium lattice to estimate $f_s<10^{-4}$ for solid helium \cite{9}, a result compatible with current measurements \cite{16}. Our dipolar supersolid does not move in a 1D configuration as in the Leggett model but has a more complex dynamics in the whole ($x,y$) plane, with both motion along the $x$ axis, where the density modulation forms, and rotation of the individual droplets. Therefore, we expect Eq.~\ref{eq4} to account only for the superfluid fraction related to the dynamics along $x$, as it does not consider the superfluidity of the individual droplets.

Because we cannot measure $\rho(x)$ experimentally, we employ numerical calculations \cite{29}. The right-hand side of Eq.~\ref{eq4} is shown in Fig.~\ref{fig3} as triangles. It drops from unity for the BEC to about 0.3 for the supersolid, a relatively large value set by the large overlap between the two central droplets (see Fig.~\ref{fig3}, inset). It then decreases for increasing $\epsilon_{dd}$, reaching almost zero at $\epsilon_{dd}$=1.5, where the droplets overlap almost vanishes. In that regime, one can recover the finite superfluid fraction of the numerical calculations by considering the droplets’ superfluidity. Indeed, applying Eq.~\ref{eq2} to the case of independent droplets and considering that each droplet’s moment of inertia about its axis is zero thanks to the cylindrical symmetry~\cite{29}, one obtains the estimate $f_s^{drop}\approx(1-\beta)/(1-\beta^2)$. Using the theoretical distributions, we get $f_s^{drop}\approx$0.5 for all the data points in the supersolid regime (black diamonds in Fig.~\ref{fig3}). This estimate is quite close to the numerical data point for $f_s$ at $\epsilon_{dd}$=1.5, and more than 2 standard deviations below the experimental data point at $\epsilon_{dd}$=1.45. Together with the qualitatively similar reduction of the two theoretical datasets for increasing $\epsilon_{dd}$, this suggests that the mechanism identified by Leggett might have a relevant role in our small dipolar supersolid. To obtain a quantitative assessment, one will need further measurements and a theoretical analysis based on a 2D analogue of the Leggett result \cite{35,36}.

We have established the superfluid nature of the dipolar supersolid by characterizing its non-classical rotational inertia. The supersolid is different from standard superfluids because of the reduced superfluid fraction thanks to the spontaneous breaking of translational invariance. The techniques we have demonstrated, with an improvement of the measurement precision and of the resolution on $\epsilon_{dd}$, will allow testing whether the superfluid fraction of the supersolid is indeed smaller than unity. Achieving larger systems might also allow studying the supersolid behavior in an annular geometry or in a 2D configuration, as well as studying the dynamics of quantized vortices in the supersolid phase ~\cite{28}.



We thank E. Lucioni for contributions to the early stages of the experiment, A. Gallemi, A. Recati, S. Roccuzzo and S. Stringari for discussions and for providing the theoretical data, D.E. Galli for discussions, A. Barbini, F. Pardini, M. Tagliaferri and M. Voliani for technical assistance. This work received funding by the EC-H2020 research and innovation program (Grant 641122 - QUIC).



\section*{Supplementary material}

\subsection{BEC and supersolid production}
The experiment starts from Bose-Einstein condensates of $^{162}$Dy atoms, with no detectable thermal fraction. The atoms are trapped in a harmonic potential created by two dipole traps crossing in the horizontal ($x,y$) plane. Typical trap frequencies are: $\omega_{x,y,z}=2\pi$(23,46,90)~s$^{-1}$. Since the trap potential is determined by the precise alignment of separate laser beams, it is subjected to slow drifts. Therefore, the trap frequencies are measured before and after each oscillation experiment and the measurement is considered valid only if the deviation is smaller than 2\%. We choose a 1:2 ratio for the trap frequencies in the ($x,y$) plane, since a small aspect ratio determines a large difference between the scissors frequencies of BEC and supersolid regimes, i.e. a relatively small $\beta$. The more elongated trap we used in previous experiments was indeed not appropriate for this type of measurements \cite{18,21}. The experimental trap frequencies are about 15\% larger than the ones used in the theoretical analysis \cite{28}. However, a homogeneous scaling of the trap frequencies does not change substantially the relative scissors frequency $\omega_{sc}^2⁄(\omega_x^2+\omega_y^2)$, which is the relevant quantity entering the moment of inertia. For the BEC regime, at the mean field level one can indeed calculate a shift around 1\% for the relative frequencies of the two trap configurations \cite{31}.

To tune the interaction parameter $\epsilon_{dd}=a_{dd}⁄a_s$, we control the contact (van der Waals) scattering length $a_s$ with magnetic Feshbach resonances, while the dipolar scattering length $a_{dd}=130 a_0$ is fixed. We employ a set of two narrow Feshbach resonances located at around 5.1~G, with widths $\Delta B_1$=32(7)~mG and $\Delta B_2$=8(3)~mG~\cite{18,37,38}. The magnetic field amplitude is calibrated by radio-frequency spectroscopy between two hyperfine states, with an uncertainty of about 1~mG \cite{18}. The resulting systematic uncertainty on the absolute value of $a_s$ is about 3~$a_0$, which corresponds to an uncertainty on $\epsilon_{dd}$ of about 3\%. Given the large experimental uncertainty on $a_s(B)$, we identify a precise $B$ to $a_s$ conversion by comparing experimental and numerical data for the critical $\epsilon_{dd}$ providing the onset of the modulated state. This calibration is repeated before and after each oscillation experiment.

The condensate is initially created at $a_s=140 a_0$, with typical atom number N=3.5×$10^4$. The scattering length is then tuned with a 70~ms ramp to $a_s=114 a_0$, close to the BEC-supersolid transition, which occurs at $a_s\approx92 a_0$ ($\epsilon_{dd}\approx1.42$). A second ramp lasting 30~ms brings the system into the supersolid regime. Typically, the supersolid lifetime is about 100~ms, preventing us from observing scissors oscillations for longer interrogation times. In the droplet crystal regime, for $\epsilon_{dd}>1.52$, the very short lifetime severely limits the accuracy of the frequency measurements. Therefore, we have excluded that regime from the present analysis.

The detection is performed by absorption imaging after a free expansion lasting $t_{exp}= 95$~ms. About 200~$\mu$s before the release of the atoms from the trapping potential, we increase the contact interaction strength by setting $a_s=140 a_0$, thus minimizing the effects of the dipolar interaction on the expansion. We record the atomic distribution in the ($x’,y’$) plane of the laboratory frame, interpreting it as a momentum-space density, $n(k_{x'},k_{y'})$. The imaging resolution is 0.2 $\mu$m$^{-1}$ (1/$e$ Gaussian width).
 
The presence of the supersolid density modulation is revealed by the characteristic side peaks in the momentum distribution. From their typical spacing $\bar{k}=$1.4$~\mu$m$^{-1}$, we deduce the presence of a single row of density maxima along the $x$ direction, with typical spacing $d=$4.5~$\mu$m. In the experiment, we do not have direct access to the exact density distribution. On the one side, a reliable in-situ imaging would require a spatial resolution comparable to the radius of the density maxima, hence smaller than 1~$\mu$m, well beyond current experimental possibilities. On the other side, modeling the initial stages of the expansion is challenging, so we cannot exactly relate $n(k_{x'},k_{y'})$ to the in-situ density distribution.

\subsection{Scissors mode excitation and analysis}
The experimental procedures employed to excite the scissors mode can also excite the axial breathing mode with lowest energy, which couples mainly to the width of the system along the $x$ direction~\cite{21}. In the supersolid regime, a too strong breathing oscillation tends to mask the supersolid behavior, shifting $\omega_{sc}$ towards the BEC value. Therefore, we employ two different methods for exciting the scissors mode in the BEC and supersolid regimes. 

For the BEC regime we efficiently excite the scissors mode by switching on temporarily (5~ms) a third optical trap intersecting the crossed dipole trap at an angle of about 0.7~rad in the ($x,y$) plane. This imprints a rotation of the atomic system in the ($x,y$) plane, with typical amplitude after free expansion of 0.3~rad. This method changes also the trap frequencies, exciting the axial breathing oscillation. The fractional amplitude of the oscillation of the $x$-width is about 20\%, as detected after free expansion. Since scissors and breathing are normal modes in the BEC, we do not expect a significant coupling between them \cite{39}. We checked experimentally the absence of a relevant coupling of the two modes: changing the amplitude of the quadrupole oscillation by a factor 4, the scissors frequency does not change. 

Crossing the BEC-supersolid transition produces naturally an axial breathing oscillation with fractional variation of the $x$-width of about 10\%~\cite{21}; the method employed for exciting the scissors mode in the supersolid regime avoids additional excitation of the breathing mode. We indeed excite the scissors mode by changing slightly the relative intensities of the two lasers producing the crossed dipole trap, for 5~ms. As the beams are not perfectly orthogonal (the relative angle is about 1.4~rad), this induces a small rotation in the ($x,y$) plane, with typical amplitude after free expansion of 50~mrad, much smaller than the deformation $\beta$. The change of trap frequencies associated to this method is negligible. By varying intentionally the amplitude of the breathing oscillation, we have checked experimentally that for oscillation amplitudes smaller than 15\% the value of the scissors frequency is unaffected also in the supersolid regime. Instead, breathing modes with amplitude larger than 15\% tend to shift $\omega_{sc}$ towards the BEC value.

For both methods, we let the system evolve in the trap for a variable time $t$, and then determine the angle $\theta'$ after the free expansion, by fitting $n(k_{x'},k_{y'})$ with the appropriate rotated 2D distribution. During the free expansion, the non-trivial change of the shape of the system can affect the evolution of the rotation angle $\theta'$ observed in the laboratory frame~\cite{40,41}. However, the scissors oscillation frequency is not affected by the free expansion, which in our range of parameters only enhances the oscillation amplitude by approximately a factor of 2~\cite{42}. 

To fit the distributions after the free expansion, we employ two different models, depending on the regime. In the BEC regime, we employ a 2D gaussian distribution; in the supersolid regime, we add a sinusoidal modulation along the direction of the lattice with periodicity $\bar{k}$ and relative amplitude $C_1$:
\begin{equation*}
\begin{split}
n(k_{x'},k_{y'})= &C_0 e^{-\frac{\left(k_{x'}\cos\theta'-k_{y'}\sin\theta' \right)^2}{2\sigma_x^2}-\frac{\left(k_{x'}\sin\theta'-k_{y'}\cos\theta' \right)^2}{2\sigma_y^2}}\\
&\left[1+C_1\cos^2(\frac{k_{x'}\cos\theta'-k_{y'}\sin\theta'}{\bar{k}}\pi+\phi)\right].
\end{split}
\end{equation*}
The evolution of the relevant fit parameter $\theta'(t)$ is then fitted with a damped sinusoid of the form:
\begin{equation*}
\theta'(t)=\theta'_0+\Delta\theta'\cos\left(\sqrt{\omega_{sc}^2-\tau^{-2}}t+\varphi\right)e^{-t/\tau},
\end{equation*}
where $\Delta\theta'$, $\theta'_0$, $\omega_{sc}$, $\tau$ and $\phi$ are fitting parameters.

\begin{table}[t]
\begin{ruledtabular}
\begin{tabular}{cc|ccc}
\hline
\multicolumn{2}{c|}{Individual datasets}	&	\multicolumn{3}{c}{Averaged data}\\
\hline
$\epsilon_{dd}$		&	$\omega_{sc}/\sqrt{\omega_x^2+\omega_y^2}$	&	$\omega_{sc}/\sqrt{\omega_x^2+\omega_y^2}$	&	$I/I_c$	&	$f_s$	\\

\hline
1.45	&	0.839$\pm$0.035	&	0.83$\pm$0.02	&	0.74$\pm$0.04	&	0.88$\pm$0.14	\\
1.45	&	0.821$\pm$0.028	&		&		&		\\
\hline
1.50	&	0.81$\pm$0.05	&	0.78$\pm$0.03	&	0.96$\pm$0.08	&	0.4$\pm$0.7	\\
1.50	&	0.75$\pm$0.04	&		&		&		\\
\hline
\end{tabular}
\end{ruledtabular}
\caption{Experimental data employed for the datapoints in the supersolid regime in Fig.~\ref{fig2} and Fig.~\ref{fig3}.}
\label{tab1}
\end{table}

\subsection{Experimental data points in the supersolid regime}
For $\epsilon_{dd}$=1.45 and $\epsilon_{dd}$=1.5, the experimental data points reported in Fig.~\ref{fig2} and Fig.~\ref{fig3} are obtained combining two sets of data acquired under similar experimental conditions on different days. In Table~\ref{tab1}, we report the measured frequencies for the individual datasets and the corresponding averaged values for the scissors frequency, the moment of inertia and the superfluid fraction. The error in the averaged data represents the standard error of the mean.

\begin{figure}[t]
\centering
\includegraphics[width=1\columnwidth]{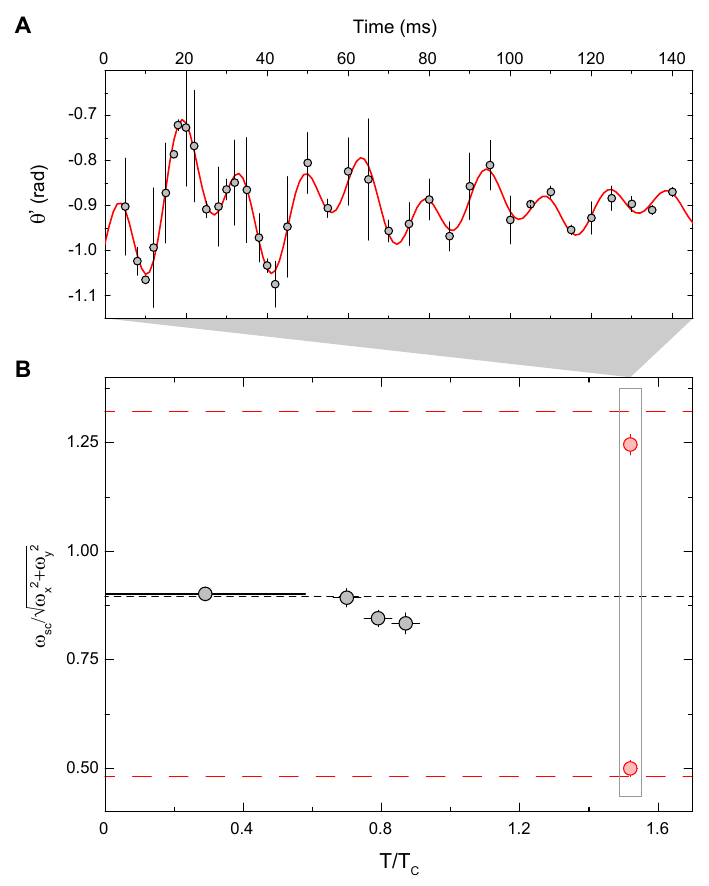}
\caption{Scissors oscillations at finite temperature. For a thermal sample in the BEC regime ($\epsilon_{dd}$=1.14, $T$=1.5~$T_c$), two oscillation frequencies are clearly visible. The angle was measured after a free expansion of 4~ms. Error bars represent the standard deviation of 4-8 measurements. B) Large temperature measurements. Measured frequencies of the scissors oscillations in the BEC regime ($	epsilon_{dd}$=1.14), for increasing temperatures. Grey dots represent condensed samples; red dots represent thermal samples. The measured frequencies for thermal sample are close to the result for weakly-interacting non-dipolar thermal gas $\omega_y\pm\omega_x$, demonstrating the collisionless nature of the system and the superfluid nature of the single-frequency oscillation of the condensed samples. We attribute the shift of the frequencies of the thermal sample to magnetostriction, while the shift of the condensed sample is due to the interaction with the coexisting thermal component.}
\label{figS1}
\end{figure}

\subsection{Finite temperature analysis}
In our system, we can safely define an equilibrium temperature only in the BEC regime. Since the minimum detectable thermal fraction in that regime is approximately 25\%, the minimum temperature we can measure is $T\approx0.6~T_c\approx$35~nK, where $T_c\approx$60~nK is the critical temperature for condensation. Since finite-temperature effects can influence the scissors mode \cite{43}, we performed a series of measurements of the scissors oscillations in the BEC regime ($\epsilon_{dd}$=1.14) at intentionally larger temperatures. To increase the temperature, we interrupt the evaporative cooling at different times. For $T<T_c$, we measure the oscillation only for the condensed component, since the thermal component after free expansion is too dilute to be detected. We determine the thermal fraction by independent measurements with shorter expansion time (25~ms). As in the low-$T$ measurements, we observe a single-frequency oscillation, but with a slightly shifted frequency. For $T>T_c$, we study instead the oscillation of the fully thermal system after 4~ms expansion, which features two distinct frequencies close to the values predicted for a weakly-interacting non-dipolar thermal gas, $\omega_{\pm}=\omega_y\pm\omega_x$, see Fig.~\ref{figS1}A~\cite{24}. Such observation demonstrates that our system is in the so-called collisionless regime, excluding instead the classical hydrodynamic behavior predicted for strongly-interacting Bose and Fermi gases \cite{24,44}. So, the single-frequency oscillation of the condensate is a direct consequence of superfluidity. A summary of these measurements is presented in Fig.~\ref{figS1}B. The frequency shift for the condensate can be justified as an effect of the interaction of the condensate with the thermal component \cite{43,45}. The shift is about 10\% close to $T_c$, and apparently becomes negligible for $T<0.7~T_c$. This suggests that the presence of a residual thermal component at the typical temperatures of the experiment is irrelevant to the dynamics of the system. We note that a similar analysis for the supersolid regime is not possible, since in our setup the supersolid can be formed only at the lowest temperatures. 

\subsection{Superfluid fraction calculation}
To calculate the upper bound on the superfluid fraction for the one-dimensional configuration, we write eq.~\ref{eq4} in the form $f_s\leq\left(\frac{1}{\lambda}\int_0^{\lambda}\frac{dx}{\rho(x)}\right)^{-1}$. Here $\rho(x)$ is the density of the system normalized to the mean density, and the integral is performed over a lattice period $\lambda$. The 1D density is obtained integrating the numerical calculated 3D density $\rho(x,y,z)$ in the $y$,$z$ directions \cite{28}. For the data points in Fig.~\ref{fig3}, we take $\lambda$ equal to the distance between the two central density maxima appearing in the supersolid phase (see inset of Fig.~\ref{fig3}). The same interval is used in the BEC phase. With this choice we neglect the contribution of the two lateral droplets, which would increase the superfluid fraction. For example, integrating between the two lateral maxima, we would find $f_s$=0.4 instead of 0.3 for the first supersolid point at $\epsilon_{dd}$=1.42. However, with this choice the superfluid fraction of the BEC becomes lower than one, since we are considering regions in which the density approaches zero due to the presence of the trap. To obtain the right value in the BEC limit, therefore, we choose to integrate between the two central density maxima.

Around $\epsilon_{dd}$=1.5, the one-dimensional upper limit is close to zero, since the overlap between the two main droplets is very small, as shown in the inset of Fig.~\ref{fig3}. In this regime, we can estimate the contribution to the superfluid fraction under rotation of the individual droplets. To do this, we employ eq.~\ref{eq2} written in the form $f_s=(1-I/I_c)/(1-\beta^2)$ and we calculate analytically $I$ and $I_c$. The Huygens-Steiner theorem allows to calculate the classical moment of inertia of the two droplets system as $I_c=2M\langle x^2+y^2\rangle=2I_0+Md^2/2$. Here, $I_0=MR^2/2$ is the intrinsic moment of inertia of each droplet, of mass $M$ and mean radius $R$, about their own center and $d/2$ is their distance from the axis of rotation of the system, $x$=$y$=0. Since each droplet is individually superfluid and has cylindrical symmetry in the ($x,y$) plane \cite{28}, $I_0$=0. Therefore, in the absence of dynamics among the droplets, the actual moment of inertia reduces to $I=Md^2/2=2M\langle x^2-y^2\rangle$. The deformation can be easily calculated as $\beta=\langle x^2-y^2\rangle/\langle x^2+y^2\rangle=I/I_c$, and one finally obtains the expression for the droplets superfluid fraction reported in the main text, $f_s^{drop}=(1-\beta)/(1-\beta^2)$. Since the radius, $R\simeq0.5~\mu$m, is much smaller than the distance, $d\simeq5~\mu$m, one finds $\beta\simeq$1 and $f_s\simeq$0.5. This simple modelling remains valid also in the presence of larger number of droplets along the x axis, as long as they have cylindrical symmetry and one can neglect the inter-droplet dynamics, i.e. the Leggett mechanism. We have therefore employed the numerical values of $\beta$ to calculate $f_s^{drop}$ for the actual theoretical distributions, which have also additional small droplets at the sides, see the inset of Fig.~\ref{fig3} and ref.~\cite{28}. We have checked that excluding the side droplets gives only a 5\% change of $f_s^{drop}$, so the two central droplets dominate the superfluid fraction.


\begin{thebibliography}{99}
\bibitem{0} A.B. Migdal, \textit{Superfluidity and the moments of inertia of nuclei.} Sov. Phys. JETP \textbf{10}, 176-185 (1970).

\bibitem{1} G. B. Hess, W. M Fairbank, \textit{Measurements of the angular momentum in superfluid helium.} Phys. Rev. Lett. \textbf{19}, 216-218 (1967).

\bibitem{2} P. J. Hakonen, O. T. Ikkala, S. T. Islander, O. V. Lounasmaa, T. K. Markkula, P. Roubeau, K. M. Saloheimo, G. E. Volovik, E. L. Andronikashvili, D. I. Garibashvili, J. S. Tsakadze, \textit{NMR experiments on rotating superfluid \textsuperscript{3}He-A: evidence for vorticity.} Phys. Rev. Lett. \textbf{48} 1838-1841 (1982).

\bibitem{3} F. Chevy, K.W. Madison, J. Dalibard, \textit{Measurement of the angular momentum of a rotating Bose-Einstein condensate.} Phys. Rev. Lett. \textbf{85}, 2223-2227 (2000).

\bibitem{4} M. W. Zwierlein, J. R. Abo-Shaeer, A. Schirotzek, C. H. Schunck, W. Ketterle, \textit{Vortices and superfluidity in a strongly interacting Fermi gas.} Nature \textbf{453}, 1047-1051 (2005). 

\bibitem{5} K. G. Lagoudakis, M. Wouters, M. Richard, A. Baas, I. Carusotto, R. Andr\'e, Le Si Dang, B. Deveaud-Pl\'edran, \textit{Quantized vortices in an exciton–polariton condensate.} Nat. Phys. \textbf{4}, 706–710 (2008). 

\bibitem{6} A. J. Leggett, \textit{Quantum liquids.} Oxford University Press, New York, ed. 1, (2006).

\bibitem{7} A. F. Andreev, I. M. Lifshitz, \textit{Quantum theory of defects in crystals.} Sov. Phys. JETP \textbf{29}, 1107–1113 (1969).

\bibitem{8} G. V. Chester, \textit{Speculations on Bose–Einstein condensation and quantum crystals.} Phys. Rev. A \textbf{2}, 256–258 (1970).

\bibitem{9} A. J. Leggett, \textit{Can a solid be “superfluid”?.} Phys. Rev. Lett. \textbf{25}, 1543-1546 (1970).

\bibitem{10} A. J. Leggett, \textit{On the superfluid fraction of an arbitrary many-body system at T=0.} J. Stat. Phys. \textbf{93}, 927-941 (1998).

\bibitem{11} M.H.W. Chan, R.B. Hallock, L. Reatto, \textit{Overview on solid \textsuperscript{4}He and the issue of supersolidity.} J. Low Temp. Phys. \textbf{172}, 317–363 (2013).

\bibitem{12} E. Kim, M.H.W. Chan, \textit{Probable observation of a supersolid helium phase.} Nature \textbf{427}, 225-227 (2004).

\bibitem{13} E. Kim, M.H.W. Chan, \textit{Observation of superflow in solid helium.} Science \textbf{305}, 1941-1943 (2004).

\bibitem{14} J. Day, J. Beamish, \textit{Low-temperature shear modulus changes in solid \textsuperscript{4}He and connection to supersolidity.} Nature \textbf{450}, 853–856 (2007).

\bibitem{15} D.Y Kim, M.H.W. Chan, \textit{Absence of supersolidity in solid helium in porous Vycor glass.} Phys. Rev. Lett. \textbf{109}, 155301 (2012).

\bibitem{16} A. Eyal, X. Mi, A.V. Talanov, J.D. Reppy, \textit{Multiple mode torsional oscillator studies and evidence for supersolidity in bulk \textsuperscript{4}He.} PNAS \textbf{113}, E3203 (2016).

\bibitem{17} J. Ny\'eki, A. Phillis, A. Ho, D. Lee, P. Coleman, J.  Parpia, B. Cowan, J. Saunders, \textit{Intertwined superfluid and density wave order in two-dimensional \textsuperscript{4}He.} Nat. Phys. \textbf{13}, 455-459 (2017).

\bibitem{18} L. Tanzi, E. Lucioni, F. Fam\`a, J. Catani, A. Fioretti, C. Gabbanini, R.N. Bisset, L. Santos, G. Modugno, \textit{Observation of a dipolar quantum gas with metastable supersolid properties.} Phys. Rev. Lett. \textbf{122}, 130405 (2019).

\bibitem{19} F. B\"ottcher, J.-N. Schmidt, M. Wenzel, J. Hertkorn, M. Guo, T. Langen, T. Pfau, \textit{Transient supersolid properties in an array of dipolar quantum droplets.} Phys. Rev. X \textbf{9} 011051 (2019).

\bibitem{20} L. Chomaz, D. Petter, P. Ilzh\"ofer, G. Natale, A. Trautmann, C. Politi, G. Durastante, R.M.W. van Bijnen, A. Patscheider, M. Sohmen, M.J. Mark, F. Ferlaino, \textit{Long-lived and transient supersolid behaviors in dipolar quantum gases.} Phys. Rev. X \textbf{9} 021012 (2019).

\bibitem{21} L. Tanzi, S.M. Roccuzzo, E. Lucioni, F. Fam\`a, A. Fioretti, C. Gabbanini, G. Modugno, A. Recati, S. Stringari, \textit{Supersolid symmetry breaking from compressional oscillations in a dipolar quantum gas.} Nature \textbf{574}, 382 (2019).

\bibitem{22} M. Guo, F. B\"ottcher, J. Hertkorn,  J.-N. Schmidt, M. Wenzel, H. P. B\"uchler, T. Langen, T. Pfau, \textit{The low-energy Goldstone mode in a trapped dipolar supersolid.} Nature \textbf{574}, 386 (2019).

\bibitem{23} G. Natale, R.M.W. van Bijnen, A. Patscheider, D.  Petter, M.J. Mark, L. Chomaz, F. Ferlaino, \textit{Excitation spectrum of a trapped dipolar supersolid and its experimental evidence.} Phys. Rev. Lett. \textbf{123}, 050402 (2019).

\bibitem{26} N. Lo Iudice, F. Palumbo, \textit{New isovector collective modes in deformed nuclei.} Phys. Rev. Lett. \textbf{41}, 1532 (1978).

\bibitem{24} D. Gu\'ery-Odelin, S. Stringari, \textit{Scissors mode and superfluidity of a trapped Bose-Einstein condensed gas.} Phys. Rev. Lett. \textbf{83}, 4452-4455 (1999).

\bibitem{25} F. Zambelli, S. Stringari, \textit{Moment of inertia and quadrupole response function of a trapped superfluid.} Phys. Rev. A \textbf{63}, 033602 (2001).

\bibitem{27} O.M. Marag\`o, S.A. Hopkins, J. Arlt, E. Hodby, G. Hechenblaikner, C. J. Foot, \textit{Observation of the scissors mode and evidence of superfluidity of a trapped Bose-Einstein condensed gas.} Phys. Rev. Lett. \textbf{84}, 2056-2019 (2000).

\bibitem{28} S.M. Roccuzzo, A. Gallem\`i, A. Recati, S. Stringari, \textit{Rotating a supersolid dipolar gas.} Phys. Rev. Lett. \textbf{124}, 045702 (2020).

\bibitem{29} See the supplemental material.

\bibitem{30} Y. Pomeau, S. Rica, \textit{Dynamics of a model of supersolid.} Phys. Rev. Lett. \textbf{72}, 2426 (1994).

\bibitem{31} R.M.W. van Bijnen, N.G. Parker, S.J.J.M.F. Kokkelmans, A.M. Martin, D.H.J. O’Dell, \textit{Collective excitation frequencies and stationary states of trapped dipolar Bose-Einstein condensates in the Thomas-Fermi regime.} Phys. Rev. A \textbf{82}, 033612 (2010).

\bibitem{32} I. Ferrier-Barbut, M. Wenzel, F. B\"ottcher, T. Langen, M. Isoard, S, Stringari, T.Pfau, \textit{Scissors mode of dipolar quantum droplets of dysprosium atoms.} Phys. Rev. Lett. \textbf{120}, 160402 (2018).

\bibitem{33} A.L. Fetter, \textit{Vortex nucleation in deformed rotating cylinders.} J. Low Temp. Phys. \textbf{16}, 533 (1974).

\bibitem{34} A.S.C. Rittner, J.D. Reppy, \textit{Observation of classical rotational inertia and nonclassical supersolid signals in solid $^4$He below 250 mK.} Phys. Rev. Lett. \textbf{97}, 165301 (2006).

\bibitem{35} C. Josserand, Y. Pomeau, S. Rica, \textit{Coexistence of ordinary elasticity and superfluidity in a model of a defect-free supersolid.} Phys. Rev. Lett. \textbf{98}, 195301 (2007).

\bibitem{36} A. Aftalion, X. Blanc, R. J. Jerrard, \textit{Nonclassical rotational inertia of a supersolid.} Phys. Rev. Lett. \textbf{99}, 135301 (2007).



\bibitem{37} E. Lucioni, L. Tanzi, A. Fregosi, J. Catani, S. Gozzini, M. Inguscio, A. Fioretti, C. Gabbanini, G. Modugno, \textit{Dysprosium dipolar Bose-Einstein condensate with broad Feshbach resonances.} Phys. Rev. A \textbf{97}, 060701(R) (2018).

\bibitem{38} F. B\"ottcher, M. Wenzel, J. N. Schmidt, M. Guo, T. Langen, I. Ferrier-Barbut, T. Pfau, R. Bombín, J. S\'anchez-Baena, J. Boronat, F. Mazzanti, \textit{Dilute dipolar quantum droplets beyond the extended Gross-Pitaevskii equation.} Phys. Rev. Research \textbf{1}, 033088 (2019).

\bibitem{39} S. Stringari, \textit{Collective excitations of a trapped Bose-condensed gas.} Phys. Rev. Lett. \textbf{77}, 2360-2363 (1996).

\bibitem{40} M. Edwards, C. W. Clark, P. Pedri, L. Pitaevskii, S. Stringari, \textit{Consequence of superfluidity on the expansion of a rotating Bose-Einstein condensate.} Phys. Rev. Lett. \textbf{88}, 070405 (2002).

\bibitem{41} G. Hechenblaikner, E. Hodby, S. A. Hopkins, O. M. Marag\`o, C. J. Foot, \textit{Direct observation of irrotational flow and evidence of superfluidity in a rotating Bose-Einstein condensate.} Phys. Rev. Lett. \textbf{88}, 070406 (2002).

\bibitem{42} M. Modugno, G. Modugno, G. Roati, C. Fort, M. Inguscio, \textit{Scissors mode of an expanding Bose-Einstein condensate.} Phys. Rev. A \textbf{67}, 023608 (2003).

\bibitem{43} O.M. Marag\`o, G. Hechenblaikner, E. Hodby, C.J. Foot, \textit{Temperature dependence of damping and frequency shifts of the scissors mode of a trapped Bose-Einstein condensate.} Phys. Rev. Lett. \textbf{86}, 3938 (2001).

\bibitem{44} M.J. Wright, S. Riedl, A. Altmeyer, C. Kohstall, E.R. S\'anchez Guajardo, J. Hecker Denschlag, R. Grimm, \textit{Finite-temperature collective dynamics of a Fermi gas in the BEC-BCS crossover.} Phys. Rev. Lett. \textbf{99}, 150403 (2007).

\bibitem{45} S. Giorgini, \textit{Collisionless dynamics of dilute Bose gases: Role of quantum and thermal fluctuations.} Phys. Rev. A \textbf{61}, 063615 (2000).

\end{thebibliography}
\end{document}